\documentclass[twocolumn]{aastex631}

\newcommand\dracoii{Draco~{\sc II}}

\usepackage[caption=off]{subfig}
\usepackage{hyperref}
\usepackage{amsmath}
\usepackage[hyphens]{xurl}

\begin{document}

\title{The JWST Resolved Stellar Populations Early Release Science Program VII.  Stress Testing the NIRCam Exposure Time Calculator}

\author[0000-0002-1445-4877]{Alessandro Savino}
\affiliation{Department of Astronomy, University of California, Berkeley, CA 94720, USA}
\email{asavino@berkeley.edu}

\author[0000-0002-5581-2896]{Mario Gennaro}
\affiliation{Space Telescope Science Institute, 3700 San Martin Drive, Baltimore, MD 21218, USA}
\affiliation{The William H. Miller {\sc III} Department of Physics \& Astronomy, Bloomberg Center for Physics and Astronomy, Johns Hopkins University, 3400 N. Charles Street, Baltimore, MD 21218, USA}

\author[0000-0001-8416-4093]{Andrew E. Dolphin}
\affiliation{Raytheon, 1151 E. Hermans Rd.,
Tucson, AZ 85756}
\affiliation{Steward Observatory, University of Arizona, 933 N. Cherry Avenue, Tucson, AZ 85719, USA}

\author[0000-0002-6442-6030]{Daniel R. Weisz}
\affiliation{Department of Astronomy, University of California, Berkeley, CA 94720, USA}

\author[0000-0001-6464-3257]{Matteo Correnti}
\affiliation{INAF Osservatorio Astronomico di Roma, Via Frascati 33, 00078, Monteporzio Catone, Rome, Italy}
\affiliation{ASI-Space Science Data Center, Via del Politecnico, I-00133, Rome, Italy}

\author[0000-0003-2861-3995]{Jay Anderson}
\affiliation{Space Telescope Science Institute, 3700 San Martin Drive, Baltimore, MD 21218, USA}

\author[0000-0002-1691-8217]{Rachael Beaton}
\affiliation{Space Telescope Science Institute, 3700 San Martin Drive, Baltimore, MD 21218, USA}
\affiliation{Department of Astrophysical Sciences, Princeton University, 4 Ivy Lane, Princeton, NJ 08544, USA}
\affiliation{The Observatories of the Carnegie Institution for Science, 813 Santa Barbara St., Pasadena, CA 91101, USA}

\author[0000-0003-4850-9589]{Martha L. Boyer}
\affiliation{Space Telescope Science Institute, 3700 San Martin Drive, Baltimore, MD 21218, USA}

\author[0000-0002-2970-7435]{Roger E. Cohen}
\affiliation{Department of Physics and Astronomy, Rutgers, the State University of New Jersey,  136 Frelinghuysen Road, Piscataway, NJ 08854, USA}

\author[0000-0003-0303-3855]{Andrew A. Cole}
\affiliation{School of Natural Sciences, University of Tasmania, Private Bag 37, Hobart, Tasmania 7001, Australia}

\author[0000-0001-7531-9815]{Meredith J. Durbin}
\affiliation{Department of Astronomy, University of California, Berkeley, CA 94720, USA}

\author[0000-0001-9061-1697]{Christopher T. Garling}
\affiliation{Department of Astronomy, The University of Virginia, 530 McCormick Road, Charlottesville, VA 22904, USA}

\author[0000-0002-7007-9725]{Marla C. Geha}
\affiliation{Department of Astronomy, Yale University, New Haven, CT 06520, USA}

\author[0000-0003-0394-8377]{Karoline M. Gilbert}
\affiliation{Space Telescope Science Institute, 3700 San Martin Drive, Baltimore, MD 21218, USA}
\affiliation{The William H. Miller {\sc III} Department of Physics \& Astronomy, Bloomberg Center for Physics and Astronomy, Johns Hopkins University, 3400 N. Charles Street, Baltimore, MD 21218, USA}

\author[0000-0001-9690-4159]{Jason Kalirai}
\affiliation{John Hopkins Applied Physics Laboratory, 11100 Johns Hopkins Road, Laurel, MD 20723, USA}

\author[0000-0002-3204-1742]{Nitya Kallivayalil}
\affiliation{Department of Astronomy, The University of Virginia, 530 McCormick Road, Charlottesville, VA 22904, USA}

\author[0000-0001-5538-2614]{Kristen B. W. McQuinn}
\affiliation{Space Telescope Science Institute, 3700 San Martin Drive, Baltimore, MD 21218, USA}
\affiliation{Department of Physics and Astronomy, Rutgers, the State University of New Jersey,  136 Frelinghuysen Road, Piscataway, NJ 08854, USA}

\author[0000-0002-8092-2077]{Max J. B. Newman}
\affiliation{Department of Physics and Astronomy, Rutgers, the State University of New Jersey,  136 Frelinghuysen Road, Piscataway, NJ 08854, USA}

\author[0000-0002-3188-2718]{Hannah Richstein}
\affiliation{Department of Astronomy, The University of Virginia, 530 McCormick Road, Charlottesville, VA 22904, USA}


\author[0000-0003-0605-8732]{Evan D. Skillman}
\affiliation{Minnesota Institute for Astrophysics, University of Minnesota, 116 Church St. SE, Minneapolis, MN 55455, USA}

\author[0000-0003-1634-4644]{Jack T. Warfield}
\affiliation{Department of Astronomy, The University of Virginia, 530 McCormick Road, Charlottesville, VA 22904, USA}

\author[0000-0002-7502-0597]{Benjamin F. Williams}
\affiliation{Department of Astronomy, University of Washington, Box 351580, U.W., Seattle, WA 98195-1580, USA}



\begin{abstract}

We empirically assess estimates from v3.0 of the JWST NIRCam Exposure Time Calculator (ETC) using observations of resolved stars in Local Group targets taken as part of the Resolved Stellar Populations Early Release Science (ERS) Program. For bright stars, we find that: (i) purely Poissonian estimates of the signal-to-noise ratio (SNR) are in good agreement between the ETC and observations, but non-ideal effects (e.g., flat field uncertainties) are the current limiting factor in the photometric precision that can be achieved; (ii) source position offsets, relative to the detector pixels, have a large impact on the ETC saturation predictions and introducing sub-pixel dithers in the observation design can improve the saturation limits by up to $\sim1$~mag. For faint stars, for which the sky dominates the error budget, we find that the choice in ETC extraction strategy (e.g., aperture size relative to point spread function size) can affect the exposure time estimates by up to a factor of 5. We provide guidelines for configuring the ETC aperture photometry to produce SNR predictions in line with the ERS data.  Finally, we quantify the effects of crowding on the SNRs over a large dynamic range in stellar density and provide guidelines for approximating the effects of crowding on SNRs predicted by the ETC.  


\end{abstract}



\section{Introduction} \label{sec:intro}
Accurately predicting exposure times is central to planning virtually any astronomical observation.  An underestimation of the exposure time results in data with signal-to-noise ratios (SNRs) that can hinder or entirely prevent science goals from being achieved. Conversely, an overestimation of the exposure time may produce better-than-expected data, but at the cost of using precious, and usually expensive, telescope resources, some of which could have been allocated toward other science programs. Exposure time calculators (ETCs) play a critical, but often underappreciated, role in balancing science objectives and efficiency. 




Among the highest profile examples is the James Webb Space Telescope (JWST) ETC.  JWST's performance is exceeding pre-launch expectations \citep[e.g.,][]{McElwain23,Rieke23,Rigby23}. The number of JWST proposals is growing rapidly each cycle, leading to record high subscription rates and increased pressure on JWST observing time.\footnote{\url{https://www.stsci.edu/contents/news/jwst/2023/jwst-observers-set-world-record-for-astronomical-proposal-submissions.html}} The allocation of JWST time, in large part, depends on the ability of the ETC to accurately model the exposure times needed to achieve a diverse range of science.  The JWST ETC is quite sophisticated and is regularly updated to reflect an ever-improving knowledge of JWST's in-flight performance.\footnote{\url{https://jwst-docs.stsci.edu/jwst-exposure-time-calculator-overview}} Nevertheless, there is a paucity of publicly available benchmarks of the ETC's performance \citep[e.g.,][]{Bagley23,Williams23}.  Though many individual science programs are able to gauge the accuracy of the ETC after they have acquired data, much of this knowledge is not recycled back to the community for future planning. 

A primary goal of the Resolved Stellar Populations Early Release Science (ERS) program (DD-ERS-1334; \citealt{Weisz23}) is to provide an empirical assessment of the NIRCam ETC relative to point source photometry performed on our data. The ERS team has been using DOLPHOT \citep{Dolphin00,Dolphin16} as the main tool for point source photometry. DOLPHOT has been used to provide well-calibrated photometry for millions of resolved stars in hundreds of galaxies and star clusters in the local Universe, often with its tailored modules for the Hubble Space Telescope (HST) and JWST \citep[e.g., ][]{Holtzman06,Dalcanton09,McQuinn10,Radburn-Smith11,Williams14,Jang17,Skillman17,Sabbi18,Williams21,Savino22,Carleton23,Riess23,Savino23,Anand24,Lee24,Peltonen24}. 

In this paper, we undertake a detailed exploration of the fidelity of the JWST NIRCam ETC predictions relative to photometric data produced by our ERS program.  Specifically, we examine a wide range of effects included in the ETC (photometric aperture properties, flat fields, saturation, etc.) and compare them to the multi-band photometric data and artificial star tests (ASTs) from our application of the DOLPHOT JWST/NIRCam module to the three ERS targets: the globular cluster M92, the ultra-faint dwarf galaxy \dracoii, and the star-forming dwarf galaxy WLM. Preliminary explorations of the NIRCam ETC are part of our program's previous papers \citep{Weisz23,Weisz24}, which generally found the ETC to be conservative relative to actual point source data acquired with JWST.  But these analyses were fairly simple.  They did not explore the range of adjustable parameters in the ETC (e.g., aperture size, filters, sky) nor did they consider the effects of crowding on the photometry. 

In this paper, we present a far more detailed analysis of the ETC's performance. We assess both the performance and limitations of the ETC relative to real observations of point sources.  We also aim to provide the community with guidance on how to bring the ETC into better agreement with observations. We limit the analysis to the NIRCam portion of the ERS data. NIRCam is the near-infrared imaging workhorse of JWST \citep[e.g.,][]{Rieke05}, and is therefore the target of our analysis. 

This paper is organized as follows. In \S~\ref{sec:Methodology}, we lay out our methodology to quantify the SNR as a function of magnitude, both for our data and for the ETC simulations; in \S~\ref{sec:Results} we compare the ETC predictions to our empirical estimates, focusing on different brightness regimes and observational conditions; and in \S~\ref{sec:Conclusions} we summarize the main results, providing guidelines to optimize observation design.

\section{Methodology}
\label{sec:Methodology}

\subsection{Photometric Catalogs}
We use photometric catalogs created as part of the ERS-1334 resolved stellar population program \citep{Weisz23, Weisz24}. The observations behind this data are thoroughly described in \citet{Weisz23} and the key exposure specifications are summarized in Table~\ref{tab:Set-up}. Likewise, the photometric reduction, performed with the software DOLPHOT \citep{Dolphin16}, has been thoroughly detailed in \citet{Weisz24}. For this work, we apply the same source selection criteria described in \citet{Warfield23}. Compared to other source selections used in our program \citep[e.g., that of][ which prioritized completeness of the sample]{McQuinn24} these cuts are focused on purity, e.g., selecting sources that are well fit by the point spread function (PSF) profile, to ensure a reliable comparison with the ETC predictions. However,  we do not cut any source based on SNR to avoid biasing the sample.

By design, each of the three targets of the survey (M92, WLM and \dracoii) are suitable for different aspects of this analysis. For instance, M92 has many high-SNR sources close to the saturation limit, WLM has a large sample of faint low-SNR sources, and \dracoii\ has a very low stellar density that is suitable to benchmark photometric performance in a very uncrowded field. Throughout this paper, we will use the most appropriate target for each regime but the conclusions are consistent within the three targets. Similarly, certain aspects of our analysis use only one filter (e.g., F090W), for illustrative purposes, but the results generally apply to all NIRCam bandpasses, as long as the PSF size and the detector pixel scale are taken into account. 

During photometric reduction, DOLPHOT estimates the SNR for each detected source. For each band, this is calculated on the basis of the total point-source counts, measured by PSF-fitting, and corrected by a local measurement of the sky brightness. The source counts are then converted into flux (counts per second) using the total exposure time across all frames in which the source is measured, and the SNR is derived by assuming Poisson statistics for photons and incorporating readout noise. Throughout this paper, we will refer to this type of SNR (e.g., purely based on source and sky Poisson statistics) as ``photon-based''.

\begin{figure}
    \subfloat
        {\includegraphics[width=0.45\textwidth]{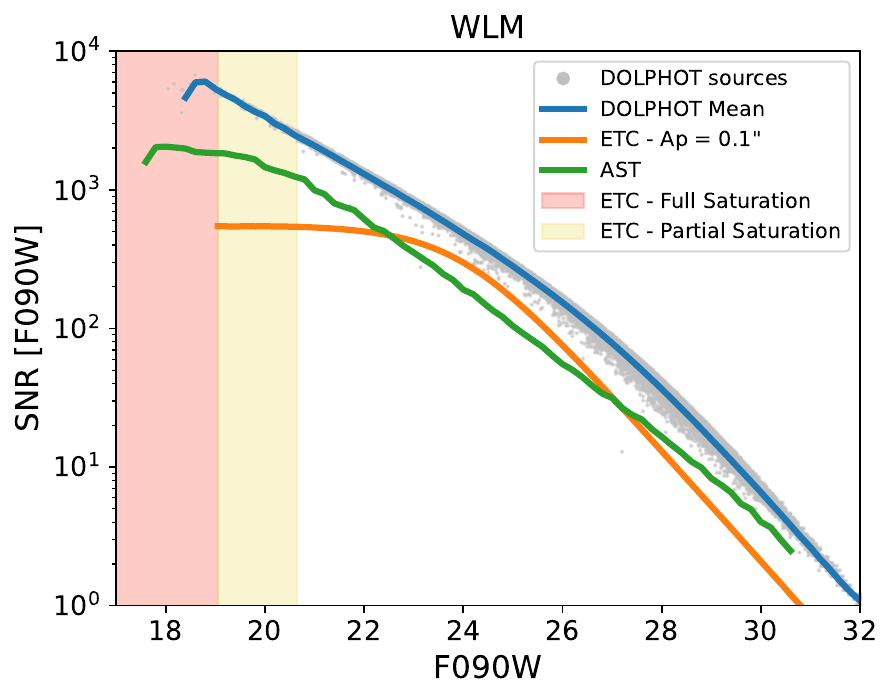}} \quad
    \subfloat
        {\includegraphics[width=0.45\textwidth]{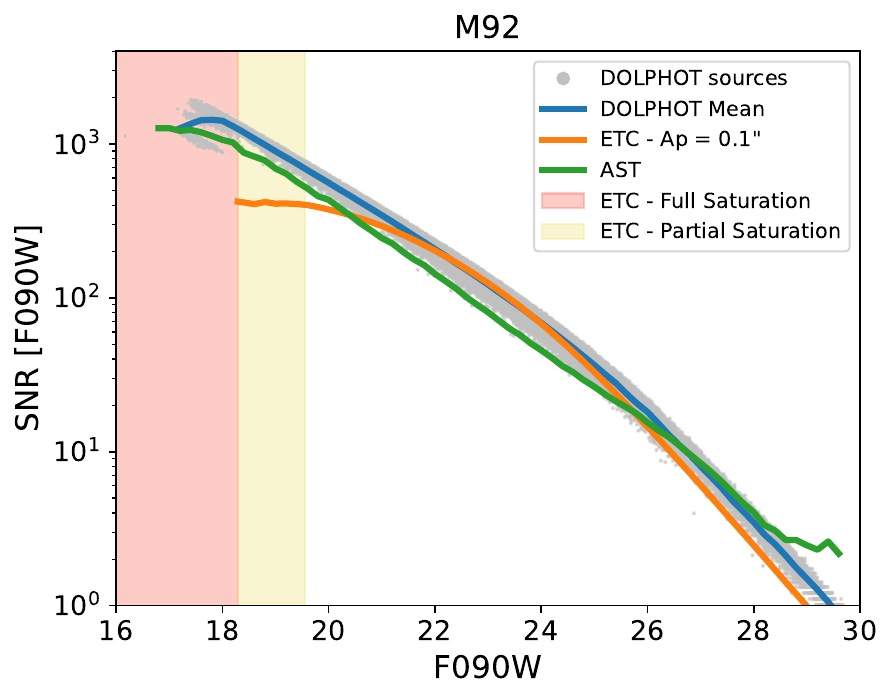}}  \quad
    \subfloat
        {\includegraphics[width=0.45\textwidth]{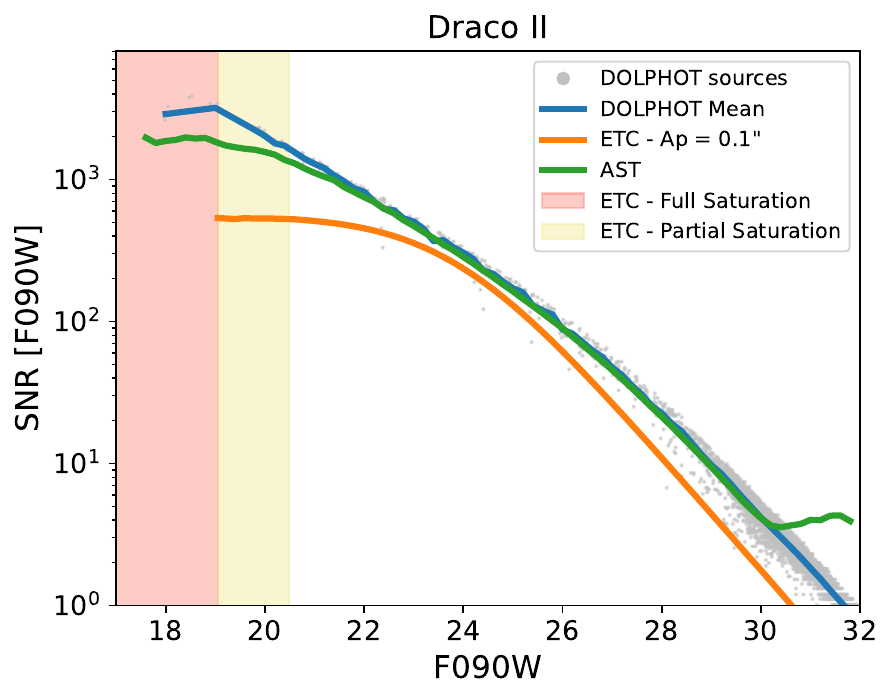}}  \quad

    \caption{SNR as a function of magnitude, for the F090W observations of WLM (top), M92 (middle), and Draco II (bottom). The grey points are the individual sources in our DOLPHOT catalog. Overlaid are the mean SNR calculated from the DOLPHOT catalog (blue line), the SNR predicted, for our observation setup, by the standard ETC configuration (orange line), and the mean SNR derived from the AST catalogs (green line). The yellow and pink shaded regions mark the partial and full saturation regimes, respectively, predicted by the ETC. }
    \label{fig:SNR}
 
\end{figure}

Figure~\ref{fig:SNR} shows the SNRs calculated by DOLPHOT, in the F090W band, for each of our targets (grey points). We use these values to calculate the $3\sigma$-clipped mean SNR of our data, in 0.2~mag bins, from the saturation magnitude, down to SNR $\sim1$ (blue lines). We calculate this mean SNR for each target and each band in our dataset. These will serve as our comparison to the SNRs estimated by the ETC.


\begin{table}[]
    \centering
    \caption{Readout pattern, number of groups, number of integrations, and number of exposures for each target and passband of our ERS NIRCam observations.}
    \begin{tabular}{lccccc}
    \toprule
    Target&Band&Readout&$N_{Group}$&$N_{Int}$&$N_{Exp}$\\
    \toprule
    M92&F090W&SHALLOW4&6&1&3\\
    M92&F150W&SHALLOW4&6&1&3\\
    M92&F277W&SHALLOW4&6&1&3\\
    M92&F444W&SHALLOW4&6&1&3\\
    \hline
    Draco II&F090W&MEDIUM8&7&4&4\\
    Draco II&F150W&MEDIUM8&7&2&4\\
    Draco II&F360M&MEDIUM8&7&2&4\\
    Draco II&F480M&MEDIUM8&7&4&4\\
    \hline
    WLM&F090W&MEDIUM8&8&9&4\\
    WLM&F150W&MEDIUM8&8&7&4\\
    WLM&F250M&MEDIUM8&8&7&4\\
    WLM&F430M&MEDIUM8&8&9&4\\
    \toprule
    \end{tabular}
    \label{tab:Set-up}
\end{table}

\vspace{10mm}
\subsubsection{On the Use of Exposure Times}
As of writing, DOLPHOT (similarly to other PSF-photometry softwares) is designed to use the same exposure time for all pixels in a frame. For JWST detectors, this time corresponds to the \textit{Time Between First and Last Measurement} of the ETC. The use of a single exposure time across the frame is correct for detectors with a single destructive read-out at the end of the exposure, such as those of HST/ACS, HST/UVIS, as well as most ground-based optical CCDs.

For the JWST detectors, which use multiple non-destructive read-outs (up-the-ramp sampling), the use of a single exposure time is generally a good assumption, except for pixels affected by partial saturation, cosmic rays, or other sources of ``jumps'' in the data. In such instances, only a portion of the ramp is used by the JWST image processing pipeline. Thus, the exposure time in that pixel is effectively reduced, which reduces the SNR. Due to its design, DOLPHOT does not account for this loss of signal and systematically overestimates the SNR in the affected pixels (this effect will be relevant in \S~\ref{sec:Sat} and \S~\ref{sec:Faint}).

Implementing pixel-dependent exposure times in the PSF-photometry algorithm is in principle possible, albeit at the cost increased computational complexity. However, this would require information on the pixel-level exposure times to be present in the JWST data products, ideally in the form of 2D exposure-time maps. While this information is computed within the JWST pipeline, as of writing, it is not explicitly provided in any of the NIRCam data products. We note that the effect of partial ramp usage is properly propagated into the image error extensions (ERR, VAR\_POISSON, VAR\_RNOISE). While this is not sufficient to reconstruct the pixel-by-pixel exposure times, algorithms solely relying on those extensions for their error accounting would correctly include this effect in their uncertainty budget.

\subsection{ETC Predictions}
\label{sec:ETC}
We compare our measurements with predictions from the v3.0 of the JWST ETC engine, Pandeia.\footnote{\url{https://jwst-docs.stsci.edu/jwst-exposure-time-calculator-overview/jwst-etc-pandeia-engine-tutorial}} We simulate NIRCam imaging of an isolated point source, with magnitudes ranging from 14 to 32, in 0.2 mag increments. We simulate observations across all targets and bandpasses, using the observational setup of our targets detailed in Table~\ref{tab:Set-up}. We also simulate an additional data point, at the magnitude corresponding to the onset of full saturation. This is determined by the ETC as the faintest magnitude for which at least one pixel reaches 95~\% of the full-well depth within the first two groups.

We use a KV5 ($T_{\rm eff} = 4250$~K; log(g)$=4.5$) source spectrum from the Phoenix stellar library \citep{Husser13} and adopt a Milky-Way extinction curve with extinction values from \citet{Schlafly11}. As we normalize the spectrum in the same bandpass in which we measure the SNR, both assumptions have an almost negligible impact. More important is the choice of an appropriate background, as it can greatly affect the photometry of faint sources. We set the Pandeia background model to the date the data were taken and at the positions corresponding to the centers of our fields of view (see \S~2 of \citealt{Rigby23} for a summary of JWST's main background components).

We calculate SNR forecasts using a range of photometric apertures, from 0.01\arcsec\ to 0.2\arcsec\ (including the ETC default of 0.1\arcsec). The inner and outer radius of the sky aperture are set to 0.22\arcsec\ and 0.4\arcsec, respectively (the ETC default). The resulting SNRs for the ETC default aperture are plotted in Figure~\ref{fig:SNR} as the orange lines.


\subsection{Optimally Weighted Extraction Predictions}
\label{sec:OWE}
We also use the output from the ETC to calculate the expected SNR using an ``optimally weighted extraction" (OWE) strategy. This is done using the prescription described in \citet{Naylor98}. We note that the Hubble Space Telescope ETC implements OWE-based SNR forecasts, alongside the aperture photometry calculations (i.e., the ``optimal'' SNR provided by the HST ETC). 

The OWE approach exploits knowledge of the instrumental PSF to measure the total source flux independently in each pixel within a given aperture. The final flux is a weighted mean across all pixels, with the weights proportional to the SNR in each pixel. On isolated sources, this approach provides comparable performance to PSF-fitting techniques.

\begin{figure*}
    \subfloat
        {\includegraphics[width=0.5\textwidth]{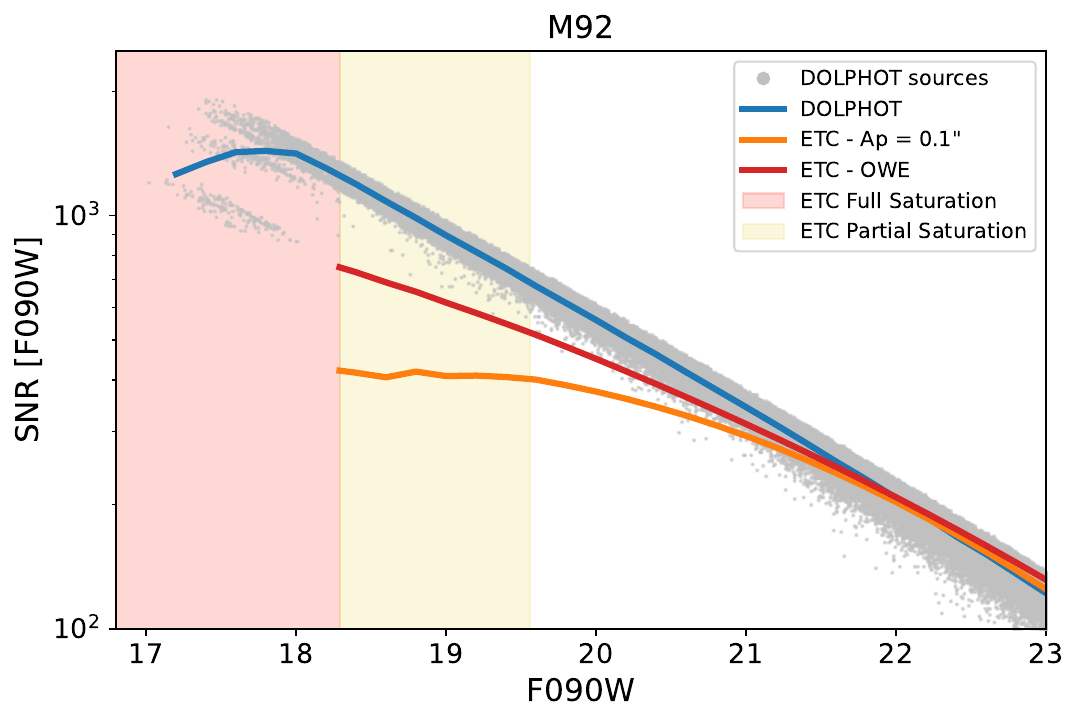}} \quad
    \subfloat
        {\includegraphics[width=0.5\textwidth]{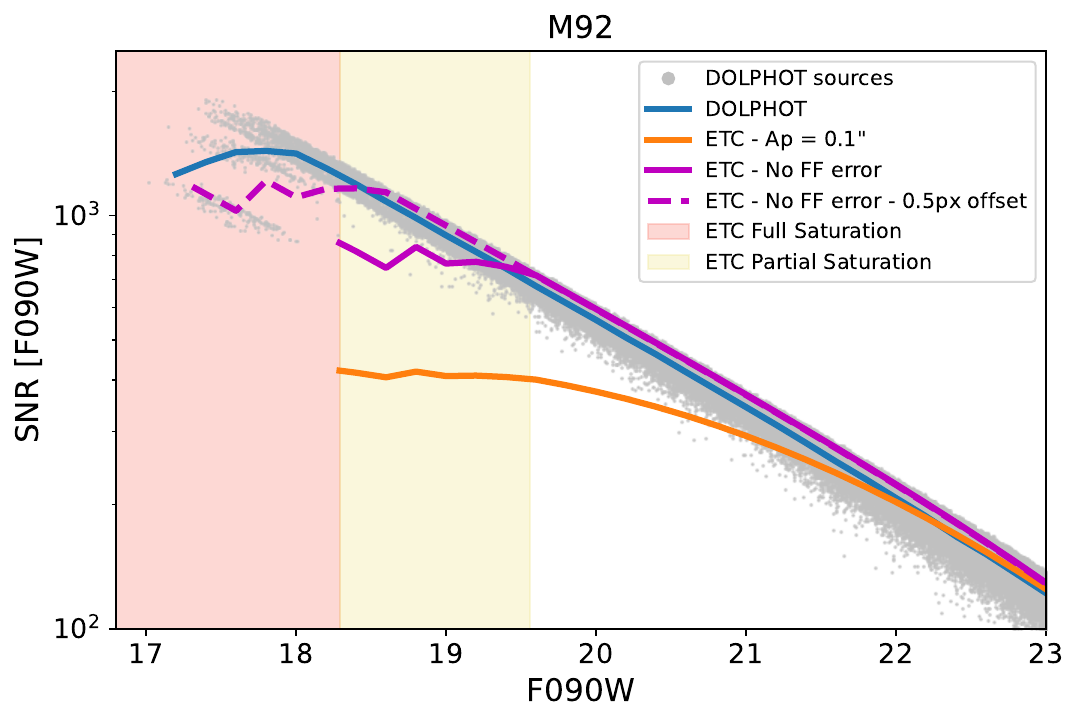}} \quad
            \caption{Mean SNR, reported by DOLPHOT, as a function of F090W magnitude for the bright sources in our M92 catalog (blue line). The grey points are the individual sources in our DOLPHOT catalog. Left: comparison with the SNR predictions of the ETC, obtained either with the standard aperture-photometry set-up (orange line) or with the OWE approach (red line). Right: DOLPHOT SNR compared to the ETC prediction with (orange line) and without (magenta solid line) flat-field (FF) uncertainties. Reported is also the SNR calculated, without flat-field uncertainties, for a point source positioned on a pixel corner (magenta dashed line). The yellow and pink shaded regions mark the partial and full saturation regimes, respectively, predicted by the ETC. }
    \label{fig:BrightEnd}
\end{figure*}

For each simulated source, we use 2D maps of the measured signal, its variance, and the value of the background flux, provided by Pandeia, to calculate the SNR using the \citet{Naylor98} formalism:
\begin{gather}
    F = \frac{\sum_{i,j}{P_{i,j}(D_{i,j}-S_{i,j})}/V_{i,j}}{\sum_{i,j}{P^2_{i,j}/V_{i,j}}}\\
    {\rm SNR}_{\rm OWE} = F \sqrt{\sum_{i,j}{P_{i,j}^2/V_{i,j}}},
\end{gather}
where $F$ is the measured flux of the source, the indices ($i,j$) are the coordinates of the discrete image pixels within a given photometric aperture, $P_{i,j}$ is the integral of the PSF over the pixel of coordinates ($i,j$), $D_{i,j}$ and $S_{i,j}$ are the total flux and the sky flux in that pixel, and $V_{i,j}$ is the the variance of the flux measurements in that pixel, which should include all relevant sources of error (e.g., Poisson noise, readout, flat-field, etc.). We use pixels within a square aperture with a side length of 7 pixels, centered on the source. We choose this size as it is comparable to the aperture size used by our DOLPHOT PSF-fitting set-up, which was determined through extensive testing in \citet{Weisz24}. The enclosed area corresponds to a circular aperture with radius 0.12\arcsec\ for the short wavelength (SW) channel and 0.25\arcsec\ for the long wavelength (LW) channel. Results from the OWE estimate are discussed in \S~\ref{sec:Bright} and \S~\ref{sec:Faint}.

\subsection{Artificial Star Tests}
All the SNR calculations described so far are for an isolated point-source. While there are regimes where this assumption is valid, such as observations of high-redshift galaxies or sparse Galactic fields, most resolved stellar fields will have some degree of crowding.  Crowding can affect the precision of photometric measurements due to flux contributions from neighboring stars.  In turn, this can alter the effective SNR.

To capture this effect, we also quantify the photometric performance of our ERS observations using the artificial star test (AST) catalogs described in \citet{Weisz24}. These tests involve injecting synthetic stars ( $\sim 3\times 10^6$ per NIRCam field) into each of the science images, running them through the same DOLPHOT photometric reduction as the real data, and comparing their known properties (e.g., input magnitudes) with what is returned from the photometric reduction (e.g., recovered magnitude).

We use the catalogs for all three targets of the ERS program (i.e., M92, WLM, and \dracoii) to assess performance in different crowding regimes. In these artificial star catalogs, we define the SNR as the inverse of the $3\sigma$-clipped standard deviation of the difference between the input and output AST magnitudes. We calculate the SNR, in 0.2~mag bins, over the whole magnitude range spanned by the AST catalog, provided that at least 100 detected ASTs are present in the bin. The resulting SNRs are showcased in Figure~\ref{fig:SNR} (green lines). From this figure, it is evident that the agreement between the AST-based SNR and the photon-based SNR of DOLPHOT varies substantially from target to target. In particular, it is important that the self-consistency of the two approaches is clearly validated in the uncrowded case of \dracoii. On the contrary, departures are observed for M92 (moderate crowding), and even more in WLM (high crowding). The effect of crowding will be discussed further in \S~\ref{sec:Crowding}.

\section{Results}
\label{sec:Results}

\subsection{The Bright End: Signal-to-Noise Comparison}
\label{sec:Bright}
We first compare measured and predicted SNRs at the bright end of our considered magnitude range. In this regime, the contributions of sky and readout to the error budget are negligible and the SNR should be purely driven by the source's signal.

The left panel of Figure~\ref{fig:BrightEnd} shows, for the bright sources in M92, the SNR measured by DOLPHOT (grey points; blue line), the ETC prediction corresponding to the default configuration (orange line; aperture photometry in a 0.1\arcsec aperture), and the OWE-based SNR (red line); as a function of F090W magnitude. While the three estimates are in good agreement for sources with F090W $\sim 23$ mag, the values quickly diverge as the source magnitude decreases. The ETC estimate plateaus at SNR $\sim 400$, the OWE SNR reaches $\sim 750$ at full saturation, and the DOLPHOT SNR steadily climbs to values over 1000.

The reason for this discrepancy is that the ETC (and, by extension, the OWE calculation) includes extra error sources which are not accounted for in DOLPHOT. In particular, the uncertainty on the NIRCam flat-field calibration is the dominant term in the high-SNR regime. We demonstrate this in the right panel of Figure~\ref{fig:BrightEnd}, which shows the effect of disabling the flat-field error treatment from the ETC predictions. For unsaturated sources, these predictions are in perfect agreement with the DOLPHOT SNRs, which assume perfect flat-fielding.

Given the multiplicative nature of the process of flat-fielding an image, the ETC treats the flat-field uncertainty as an extra error term, proportional to the flux in a pixel. This term limits the theoretical SNR that can be achieved. The current implementation of Pandeia assumes a signal of 18,000 $e^-/px$ in the NIRCam master flat. This corresponds to a 0.75\% uncertainty in the absolute flux and a maximum SNR of 133 per pixel, per dither. The maximum SNR that can be achieved per source is higher than this value, because multiple pixels in the aperture are used to calculate photometry and because our setup uses several independent exposures.


We note that the value of 18,000 $e^-/px$ (i.e., the 0.75\% flux uncertainty per pixel), used by the ETC, is determined by pre-launch flat-field measurements. Current in-flight flat-field calibration files (context jwst\_1230.pmap) have comparable uncertainties. Across the 8 NIRCam wide filters, and across all detectors, the median flat field uncertainty is 0.73\%. There is however, meaningful scatter in the precision of current flat-field calibrations, with filter/detector median uncertainties that range from 0.5\% (F277W, chip A5) to 1.6\% (F090W, chip A3). Calibrations for the SW filters on module A, in particular, have typical uncertainties above 1\%.

This means that flat-field uncertainties can be somewhat underrepresented or overrepresented in the current ETC implementation, depending on the filter and detectors 
 of the observations. Concerned observers should consult the uncertainties in the relevant flat-field reference files (and how they compare to the ETC baseline of 0.75\%) to gauge the size of this effect. We expect that, as JWST operations progress, updated calibration files will both reduce the overall size of flat-field uncertainties and improve homogeneity across filters and detectors, as much as permitted by other detector systematic effects, such as $1/f$ noise.

In practice, the precise SNR value achieved for the brightest sources is seldom a determining factor in program design. Specific science cases, for which exquisite photometric accuracy is required, might be better suited for NIRCam observing modes explicitly designed to guarantee high photometric stability. Differential photometry through the NIRCam Time Series Imaging mode is an example of such instances.

However, in cases where NIRCam standard Imaging mode is used, and high photometric accuracy is necessary, a few strategies can be helpful to minimize the impact of the flat-field calibration uncertainty. First, using multiple \textit{dithered} exposures increases the maximum achievable SNR by the square root of the number of exposures. Dithering in this case is critical, as the SNR gain is only achieved by observing a source in different regions of the detector.\footnote{Standard sub-pixel dithers are sufficiently large to provide mitigation of flat-field uncertainties. Small-grid dithers, on the other hand, use smaller pointing offsets and are less effective, especially for the LW channel.}

Second, where possible, performing photometry over a large aperture increases the number of pixels that contribute to the photometric measurement and increases the maximum SNR. However, because most of the source flux is contained within a small number of central pixels, the gain is relatively modest. For the SW channel, increasing the extraction aperture from 0.1\arcsec\ to 0.5\arcsec\ increases the SNR by roughly 25\%. The gain for the LW channel is somewhat larger, due to the larger PSF size, and ranges from 30\% to 45\%.

\begin{figure*}
 \subfloat
        {\includegraphics[width=0.45\textwidth]{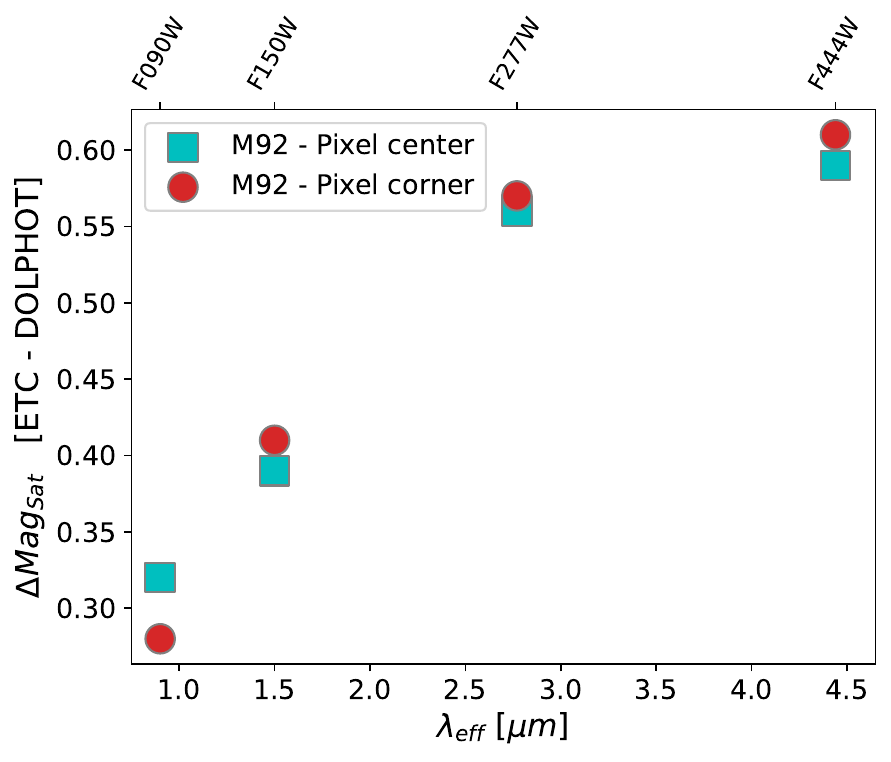}}\quad
    \subfloat
        {\includegraphics[width=0.45\textwidth]{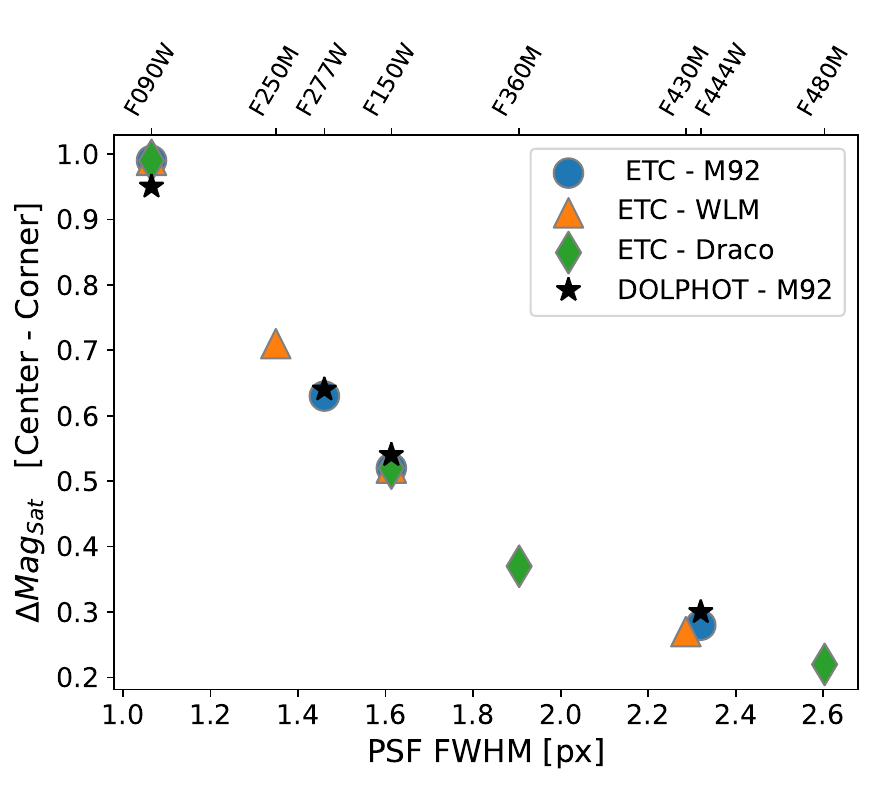}}

    \caption{Illustrative comparison between predicted and observed saturation magnitudes. Left: Difference between the full-saturation magnitude predicted, for our M92 setup, by the ETC and the values inferred from our catalog, as a function of filter effective wavelength. Cyan squares refer to sources lying on the pixel center while red circles refer to sources lying on a pixel corner. Due to the limitations described in the text, care needs to be taken when extending our results to observation design. Right: Difference between the saturation magnitude of sources centered on a pixel and sources that fall on a pixel corner, as a function of PSF FWHM, in pixels. The ETC predictions are shown for the M92 (blue circles), WLM (orange triangles), and \dracoii\ (green diamonds) set ups. The values inferred from our M92 catalog are reported as black stars. For each data point, the corresponding filter is reported on the top axis.}
    \label{fig:SatMag}
 
\end{figure*}

Finally, if reliable PSF models are available \citep[which is the case for NIRCam with WebbPSF v1.2 and later, e.g., ][]{Weisz24}, PSF-fitting or OWE photometry can greatly mitigate the effect of flat-field uncertainties. This is clearly shown in Figure~\ref{fig:BrightEnd}, with OWE photometry reaching almost double the SNR, compared to aperture photometry, at the onset of full saturation. This is due to the way pixel-based measurements are weighted in the OWE approach. As the variance in the central pixels increase, due to the proportional nature of flat-field uncertainties, the weights are adjusted so that information is more evenly extracted from the pixels within the aperture, and the maximum SNR is therefore increased. This, however, is only the case if the flat-field uncertainty is explicitly accounted for in the pixel-level variance.

On the other hand, while PSF-based techniques can help reduce the uncertainties related to flat-fielding, they also introduce a photometric precision ceiling, dictated by the accuracy of the PSF model itself. As discussed in \citet{Weisz24}, current NIRCam PSF models are in very good agreement with observations. However, the residual differences can still be relevant, if very high precision (SNR$\,\gtrsim200-1000$, depending on the filter) is desired. In such cases, the observer should carefully evaluate the uncertainty budget associated with both aperture photometry and PSF-fitting, to determine which strategy offers better performance.

\subsection{The Bright End: the Saturation Regime}
\label{sec:Sat}
For observations of very bright sources, or where a large dynamic range is required, an accurate modeling of saturation is a key element in observation design. We therefore compare the ETC-based saturation magnitudes to the properties of bright sources in our M92 catalog.

Pixels on JWST detectors are characterized by two distinct saturation levels. Partial saturation happens when a pixel saturates within a ramp, but after the end of the second group.\footnote{For an overview of the different elements that make up a NIRCam exposure, see \url{https://jwst-docs.stsci.edu/jwst-near-infrared-camera/nircam-instrumentation/nircam-detector-overview/nircam-detector-readout-patterns}.} In this case, flux can still be measured from the unsaturated portion of the ramp, but the corresponding reduction in exposure time will result in a reduction in SNR. This is shown in Figure~\ref{fig:BrightEnd}, where the SNR of the ETC prediction, when flat-field errors are ignored, flattens as soon as partial saturation in the central PSF pixel onsets (right panel, magenta solid line). As the source gets brighter, saturation in the central pixel eventually occurs during the first two groups. In this case, a flux measurement is not possible and the source is considered to be fully saturated.\footnote{In principle, the JWST pipeline can use the first frame of the first group, named Frame 0, to measure the flux of bright sources and decrease the full saturation magnitude. However, as such capability is not yet fully calibrated, it is not used by default, and it is not implemented in the ETC.}

\begin{figure*}
\centering
    \subfloat
        {\includegraphics[width=0.9\textwidth]{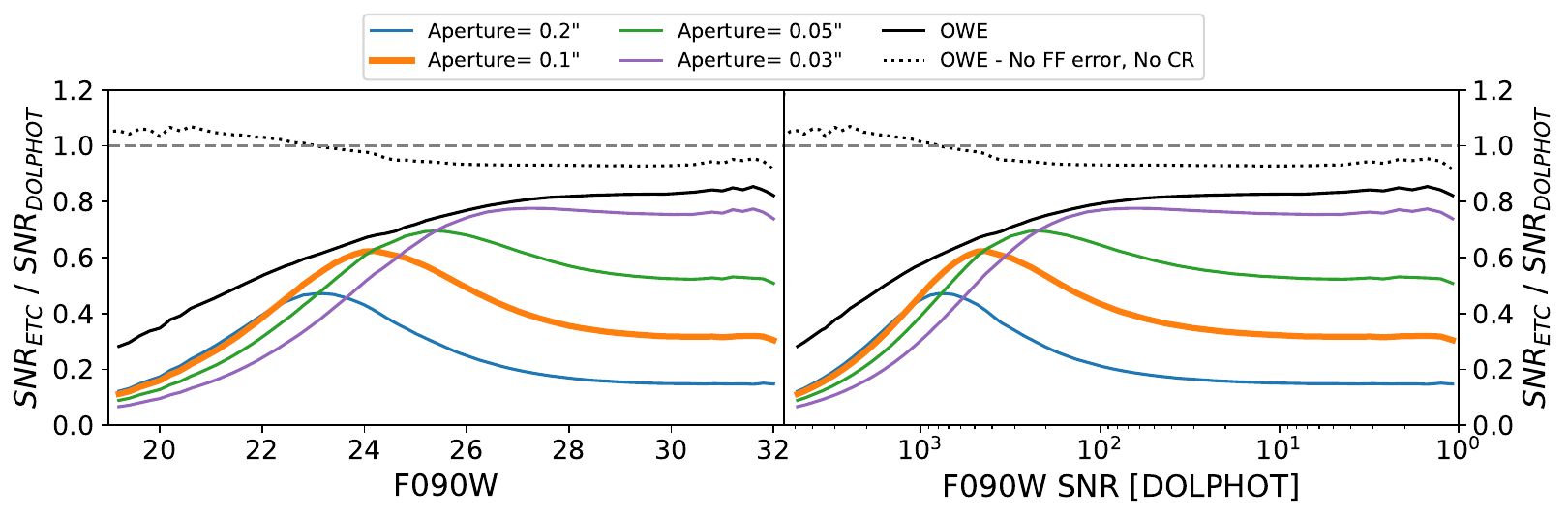}} \quad

    \subfloat
        {\includegraphics[width=0.9\textwidth]{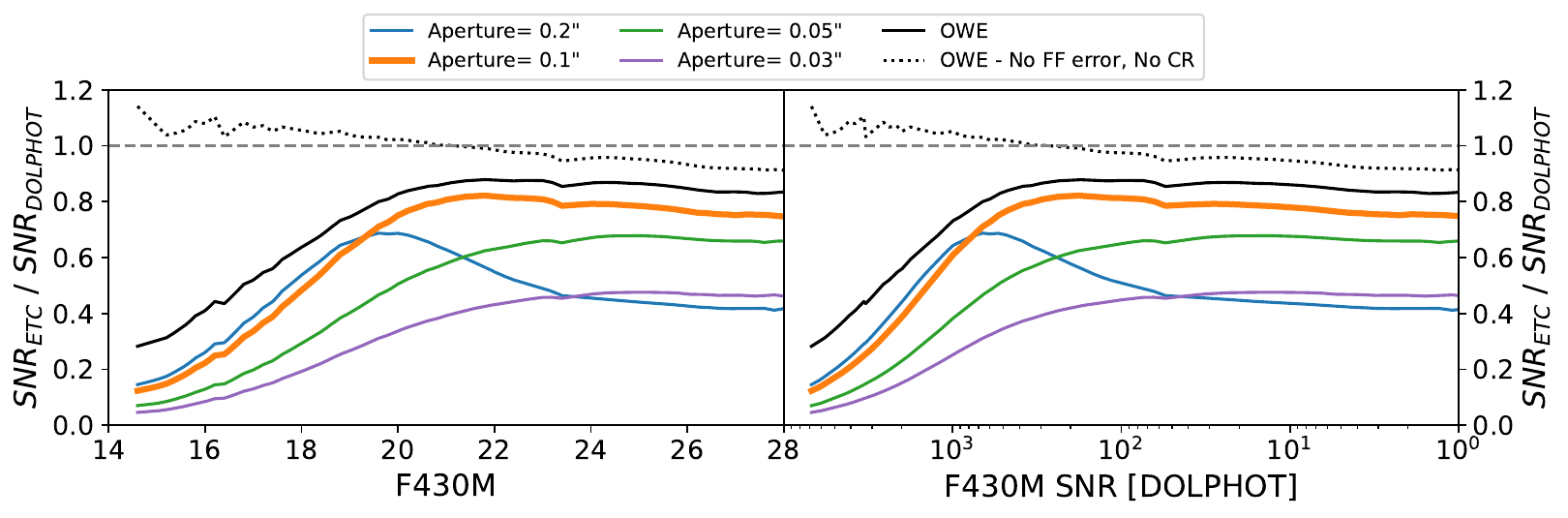}}  \quad

    \caption{Ratio between the ETC SNR prediction and the DOLPHOT SNR, for WLM, as a function of magnitude (left panels) and SNR (right panels). The comparison is illustrated for the F090W (top panels) and F430M (bottom panels) filters. Different ETC apertures are marked by different colors. The default ETC aperture is highlighted as the thick orange line. The OWE calculations are also reported (solid black line). The dotted black line is calculated using OWE SNR, without flat-field and cosmic ray error terms.}
    \label{fig:Ratios}
 
\end{figure*}

At present, DOLPHOT has no information about pixel partial saturation (e.g., pixel-level exposure-time maps). The DOLPHOT SNR for partially saturated sources is calculated using the same total exposure time as in the case of non-saturated sources, and it is therefore overestimated. This also means that we cannot use our catalogs to benchmark the ETC partial saturation forecasts.

Interestingly, for the brightest sources in our DOLPHOT catalog, there is a break in the SNR profile (at $F090W\lesssim 18$). However, this is not due to partial saturation. The break in the DOLPHOT SNR appears because certain bright stars are only marked as fully saturated in a subset of the ERS exposures, while they are well photometered in the remaining exposures. This allows the measurement of a meaningful magnitude but reduces the SNR, due to the smaller number of exposures used. These stars can be identified in Figure~\ref{fig:BrightEnd} as the two separate sequences at very bright magnitudes (F090W$\lesssim18$), just below the main SNR sequence. The fact that certain stars only saturate in specific exposures is due to the effect that dithering has on source positions within the image, as we detail below.

For a given read-out pattern, the precise magnitude of saturation onset, both full and partial, depends on the source position within the central PSF pixel. The ETC default setting is to place the source in the pixel center. This results in the most conservative saturation magnitude, as most of the source's flux is concentrated in the central pixel. As the source centroid moves towards the corner of the central pixel, the flux gets more evenly distributed over the neighboring pixels, resulting in a brighter saturation magnitude. This is illustrated in the right panel of Figure~\ref{fig:BrightEnd}, where we show that a source with offset of 0.5 px, in both X and Y, reaches partial/full saturation at significantly brighter magnitudes (magenta dashed line). Therefore, for dithered observations, a magnitude range exists in which stars only saturate in the frames in which the star centroid is sufficiently close to the pixel center.

To measure saturation magnitudes in our ERS catalogs that can be meaningfully compared to the ETC predictions, we make the assumption that the brightest observed star with at least one good photometric measurement marks the saturation magnitude for stars falling on a pixel corner. Stars brighter than this limit cannot be photometered with DOLPHOT, regardless of their position in the frame. This approach is reasonable, provided that enough bright stars exist in the frame to efficiently sample the magnitude/pixel-position parameter space. Among our targets, only M92 is therefore suitable for this exercise. Measuring the saturation magnitude for perfectly centered stars is somewhat trickier. In principle, it would be the magnitude at which bright stars begin to be discarded in at least one exposure. However, many other effects can result in bright stars being discarded, such as crowding, cosmic rays or warm pixels. For this reason, we visually inspected all bright stars, in each band, that were discarded in at least one exposure. We identify the faintest isolated star that shows signs of saturation in its core and mark its magnitude as our saturation limit. \footnote{In our images, we find that the data quality flag of 2 (saturated pixel) is only effective for tracking stars much brighter than the expected saturation threshold. For stars closer to the saturation limit, we find the typical signature to be the presence of NaN values in the source core (in the SCI extension), accompanied by a ring of pixels with data quality flag of 4 (jump detected), around the source core.}

The left panel of Figure~\ref{fig:SatMag} shows how the ERS saturation limits compare to the ETC estimates, for the four M92 filters. We find reasonable consistency, between 0.3 and 0.6 mag, between the ETC-based saturation limit and what was measured in our M92 catalog. This is the case for both perfectly centered sources and offset sources. The ETC forecast appears to be somewhat conservative, especially for redder filters. This could be partially due to the use, in the ETC, of PSF models that do not include broadening by detector effects (WebbPSF v1.1) and overestimate the flux in the central pixel. However, it is unclear if this effect alone can explain the magnitude of this difference. Furthermore, our analysis is hampered by the aforementioned difficulties of measuring an unambiguous saturation magnitude and by the limited scope of our experiment (only one target, no accounting of chip-to-chip differences, etc.). Therefore, we advise caution in taking our results at face value and recommend that, if photometry of very bright sources is a priority, the official ETC saturation limit be used in the observation design.

As discussed above, dithered observations place bright sources on different pixel positions, which can relax the saturation limit.\footnote{Even for undithered observations, some bright stars will always be detected above the saturation limit, due to the random placement of their centroid. Completeness, however, will be greatly reduced.} Due to the efficient design of NIRCam sub-pixel dithers \citep{Anderson09}, a modest amount of dithers (4-6) should be sufficient to ensure that most sources fall close to a pixel corner in at least one frame. A higher number of dithers could be desired if bright star completeness is a priority but, in such cases, the use of a shallower read-out pattern is likely to be a more efficient solution.

The right panel of Figure~\ref{fig:SatMag}, shows the change in the full saturation magnitude that can be achieved by shifting the source centroid from the center to the corner of a pixel, for the 12 observation set-ups of Table~\ref{tab:Set-up}, based on ETC predictions. The size of this effect strongly depends on the FWHM, in pixels, of the PSF. For filters with narrow PSFs, such as F090W, the saturation magnitude for off-center sources can decrease by as much as 1 mag, while very broad PSFs, such as that of the F480M band, only result in a modest change of 0.2 mag. For the four filters present in our M92 dataset, this prediction is in great agreement with our empirical measurements, supporting the consideration of this effect in the design of bright star observations.

\begin{figure}
        \includegraphics[width=0.45\textwidth]{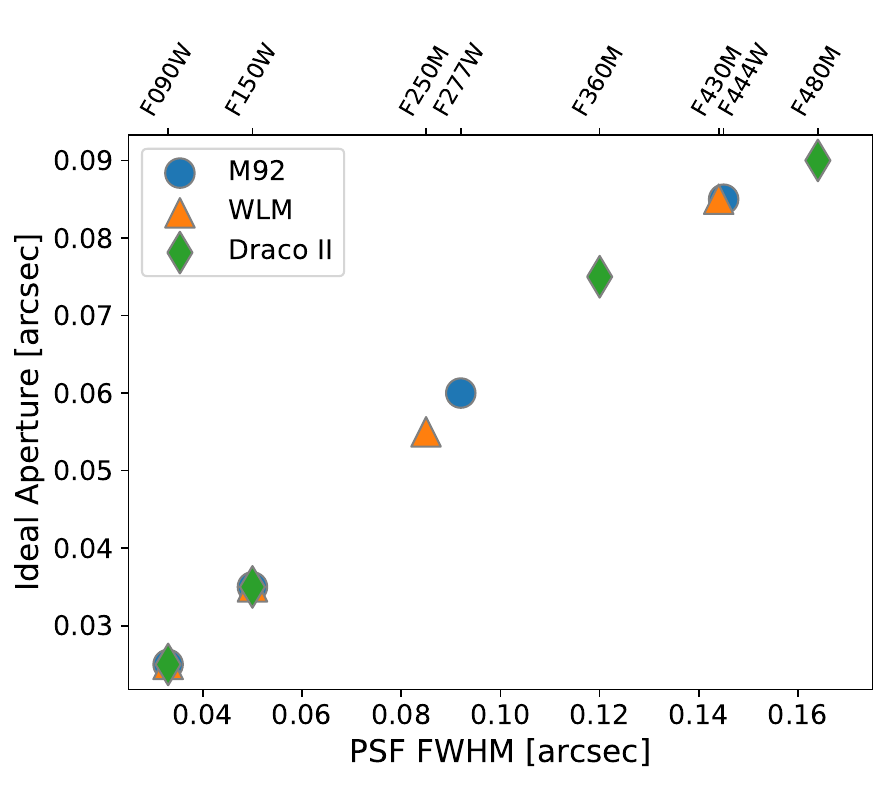}

    \caption{The photometric aperture radius that maximizes the SNR reported by the ETC for faint sources (SNR $<10$), as a function of PSF FWHM, in arcsec. The blue circles refer to the M92 setup, the orange triangles to the WLM setup, and the green diamonds to the \dracoii\ setup. For each data point, the corresponding filter is reported on the top axis.}
    \label{fig:BestAp}
 
\end{figure}

\subsection{The Faint End: the Impact of the Extraction Strategy}
\label{sec:Faint}
Accurately estimating the SNR achieved for faint sources is often a major component of observation design. Many observing programs, including our ERS program, are constructed to image sources down to a given limiting magnitude, or to achieve a target SNR at a specific magnitude. This in turns drives the overall time allocation request.

Figure~\ref{fig:Ratios} shows the ratio between the DOLPHOT-based SNR in WLM's point sources and a selection of the ETC-based predictions (described in \S~\ref{sec:ETC} and \S~\ref{sec:OWE}), for the F090W and F430M filters, as a function of both magnitude and DOLPHOT SNR. We plot SNRs based on both aperture photometry and the OWE approach (the latter will be discussed later in this section).  There are several major takeaways from this figure. First, the effect of flat-field uncertainties (see \S~\ref{sec:Bright}) at bright magnitudes (SNR $\gtrsim 400$) is obvious, as the ETC-based SNRs are much lower than the DOLPHOT estimate, which assumes perfect flat-fielding. This effect becomes subdominant, but still appreciable, at fainter magnitudes, because other uncertainties (photon statistics, sky, read-out noise, all included in DOLPHOT) dominate the error budget.

\begin{table*}[t]
    \centering
    \caption{For each target and filter, we report the PSF FWHM; the ideal aperture radius for aperture photometry ($IdAp$); the SNR increase for faint sources, when using either the ideal aperture ($SNR_{IdAp}/SNR_{Def}$) or the OWE approach ($SNR_{OWE}/SNR_{Def}$) instead of the default ETC configuration; the magnitude at which SNR=10 is achieved, when simulating our observations with the default ETC configuration ($m_{SNR=10}$); and the fraction of the exposure time needed to reach SNR=10 at that magnitude, by using either the ideal aperture ($t_{IdAp}/t_{Def}$) or the OWE approach($t_{OWE}/t_{Def}$).}
    \begin{tabular}{llccccccc}
    \toprule
    Filter&Target&FWHM&$IdAp$&$SNR_{IdAp}/SNR_{Def}$&$SNR_{OWE}/SNR_{Def}$&$m_{SNR=10}$&$t_{IdAp}/t_{Def}$&$t_{OWE}/t_{Def}$\\
    \toprule
    F090W&M92&0.033\arcsec&0.025\arcsec&2.16&2.23&26.43&0.52&0.49\\
    F150W&M92&0.050\arcsec&0.035\arcsec&1.78&1.86&26.11&0.51&0.47\\
    F277W&M92&0.092\arcsec&0.060\arcsec&1.21&1.28&25.34&0.77&0.69\\
    F444W&M92&0.145\arcsec&0.085\arcsec&1.01&1.11&24.07&0.97&0.81\\
    \hline
    F090W&Draco II&0.033\arcsec&0.025\arcsec&2.39&2.46&28.10&0.22&0.21\\
    F150W&Draco II&0.050\arcsec&0.035\arcsec&1.84&1.93&27.38&0.35&0.32\\
    F360M&Draco II&0.120\arcsec&0.075\arcsec&1.06&1.13&25.43&0.93&0.81\\
    F480M&Draco II&0.164\arcsec&0.090\arcsec&1.00&1.11&23.95&1.00&0.81\\
    \hline  
    F090W&WLM&0.033\arcsec&0.025\arcsec&2.48&2.55&28.29&0.20&0.19\\
    F150W&WLM&0.050\arcsec&0.035\arcsec&1.92&1.99&27.85&0.32&0.30\\
    F250M&WLM&0.085\arcsec&0.055\arcsec&1.35&1.40&26.27&0.56&0.52\\
    F430M&WLM&0.144\arcsec&0.085\arcsec&1.02&1.10&25.06&0.96&0.83\\

    \toprule
    \end{tabular}
    \label{tab:SNRBoost}
\end{table*}

Second, in the low-SNR regime, the ETC SNR forecasts are lower than DOLPHOT-based SNRs.  Moreover, the aperture-photometry SNR has a strong dependence on the extraction strategy. For instance, for a point source in WLM with F090W $=28$, DOLPHOT reports a mean SNR of 36.6 at the same magnitude.  With the same WLM observing set up, the ETC predicts a SNR
 of 13.0 using aperture photometry and the default ETC aperture of 0.1\arcsec. Using a 0.03\arcsec aperture results in SNR of 28.20. The smaller aperture provides a SNR that is 2.17 times larger and, if used in the observation design, would result in a reduction in exposure time by approximately a factor of 5. In the case of the F430M filter, for a source with F430M $=24$, DOLPHOT reports an SNR of 32.5, compared to an ETC-based SNR of 25.7 for the default ETC set up, and an SNR of 15.3 for a 0.03\arcsec\ aperture. In this case, the default ETC aperture performs better than the smaller aperture.

The reason why the extraction aperture is so critical in determining the SNR of faint sources is due to the importance of sky uncertainties in the error budget of these sources. The light profile of a point source rapidly falls off with distance from the PSF centroid, while the local sky brightness is roughly constant. Therefore, as the pixel distance from the source centroid increases, the sky contribution to the total counts in that pixel steadily increases and, in the case of a faint source, rapidly becomes dominant.

When using a large photometric aperture, the increased uncertainty due to the inclusion of many sky-dominated pixels outweighs the small gain in signal provided by those pixels, resulting in a lower SNR. As the aperture decreases, the sky error term also decreases, resulting in higher SNRs. Eventually, the aperture is so small that a significant fraction of the source light is excluded, and the SNR decreases again.

This means that, for a given PSF, source magnitude, and sky brightness, there is an ideal aperture that maximizes the extracted SNR. If performing photometry over such an aperture is feasible (e.g., the source is isolated enough), configuring the ETC to use that aperture value will result in maximum program design efficiency. Such an aperture will also provide a rough approximation of the SNR obtainable through PSF-fitting photometry.

Figure~\ref{fig:BestAp} reports the ETC aperture value that maximizes the SNR estimate of faint (SNR $<10$) point sources for our ERS observing set ups, as a function of PSF FWHM. The ideal aperture radius is reported in Table~\ref{tab:SNRBoost} and ranges from 0.025\arcsec to 0.09\arcsec. Figure~\ref{fig:BestAp} provides a good rule of thumb to configure the ETC for increased efficiency. However, these values have been obtained for the specific set up of our observations (Table~\ref{tab:Set-up}), on relatively faint sources (e.g., $27\leq F090W\leq 31$), and using the sky model tailored to our observations (i.e., specific target coordinates and observation date). Through limited testing, we conclude that the ideal aperture size, for the photometry of faint sources, seems to have a relatively weak dependence on the specific observing conditions (exposure time, source magnitude, sky brightness). However, in the cases of significantly different observing set ups, such as short observations of bright sources, the sky error contribution might be substantially different and we encourage experimentation with the ETC to identify the most suitable aperture. We also note that, as of this writing, the ETC employs PSF models from WebbPSF v1.1, which do not include broadening from detector effects \citep[e.g.,][]{Plazas18}. Implementation of updated, broader PSF models (e.g., WebbPSF $>$ v1.2), should slightly increase the ideal aperture radius as a function of filter.

Adopting the ideal photometric aperture can greatly improve the reliability of SNR forecasts and, in the current ETC implementation, it is the most effective way to boost the efficiency of faint point-source observations. However, this strategy still presents two limitations. First, if the real photometry is obtained through PSF-fitting, or analogous techniques, even the best aperture photometry set up is expected to somewhat underestimate the SNR forecast. Second, the dependence of the ideal aperture size on filter, sky brightness, and source magnitude adds iterations and complexity to the observation design process, which, in the case of JWST, already involves the optimization of many exposure-related parameters.

One option to address both issues would be the addition of an OWE-based SNR output to the ETC products. By exploiting knowledge of the PSF to maximize the information extracted from each pixel, OWE photometry has been shown to provide similar performance, on isolated point sources, to PSF-fitting \citep[e.g.,][]{Naylor98}. Furthermore, the OWE approach provides optimal SNR estimates over all filters and observing set ups, without the need to change the extraction parameters. This can be appreciated in Figure~\ref{fig:Ratios}, where we show that the OWE-based SNRs of faint sources (solid black line) are the closest to the empirical SNRs for both F090W and F430M. The use of OWE-based SNRs, therefore, simultaneously provides a better estimation of PSF-fitting performance and simplifies observation design.

As long as the ETC extraction aperture is set to the ideal radius, the increase in SNR provided by the OWE is relatively modest, between 3\% and 10\%, translating into an exposure time reduction between 5\% and 20\%. The difference is much more dramatic in comparison to the default ETC configuration (aperture of 0.1\arcsec). Table~\ref{tab:SNRBoost} shows the increase in SNR obtained through the use of either the ideal aperture-photometry radius or the OWE method, over the default ETC setting. We also report the reduction in exposure time, had these strategies been used in our ERS observation design. For the SW channel, using OWE-based SNRs over the default ETC setting would have decreased the exposure time by as much as a factor of 5 in F090W (a factor of 3 in F150W). The gains for the LW channel are smaller, but still lead to a 15-50\% reduction in exposure time. Generally, the need for OWE to properly approximate PSF-photometry performance is greater for bluer filters and for fainter sources. However, the resulting improvements in observing efficiency are significant over the whole NIRCam spectrum and for a large range of source magnitudes.

An additional takeaway from Figure~\ref{fig:Ratios} is that even OWE-based SNRs in WLM seem to only achieve $\sim 85\%$ of the SNR reported by DOLPHOT for faint sources. However, we argue that this difference is mostly due to non-ideal effects that are not accounted for in the photon-based SNR of DOLPHOT. While less important in the low-SNR regime, the effect of current flat-field uncertainties is still non-negligible for faint sources. Even more relevant is the effect of cosmic rays on the SNR prediction. Signal loss due to cosmic rays is effectively modeled, in the ETC, as a uniform reduction in exposure time, across all pixels. This is intended to provide a statistical approximation for the effect of cosmic rays, which is stochastic in nature. The resultant impact on the SNR is more important for faint, sky-dominated sources and, in the case of our WLM set-up, is on the order of 10-15\%.

The effect of cosmic rays on the source SNR is only accounted for by DOLPHOT for those pixels that experienced a cosmic ray event during the first two groups, which causes a total loss of signal. The majority of cosmic ray events, however, will occur in different parts of the ramp. This results in a reduction of signal, due to part of the ramp being discarded, and cannot be accounted for by DOLPHOT without pixel-level exposure time information. Indeed, turning off both flat-field uncertainties and cosmic ray noise, in Figure~\ref{fig:Ratios}, results in OWE SNRs that are consistent with the DOLPHOT estimate within 7\%. This suggests that part of the difference between the OWE estimate (solid black line) and the DOLPHOT SNR is due to current limitations in the DOLPHOT error budget, and that adding OWE-based SNRs to the ETC should provide a very good approximation of PSF-fitting performance. The remaining discrepancy between our data and the ETC likely arises from the imperfect match between the DOLPHOT and OWE extraction procedure, and relates to factors such as the precise geometry of the aperture, the size and location of the sky aperture, deviations from Poisson statistics, and, as we discuss below, the value of the local background flux.

\subsection{The Faint End: the Impact of the Sky}


Another major element required to accurately estimate the SNR of faint source observations with JWST is a reliable prediction of the local sky brightness. The JWST ETC uses a background model that includes several components, including zodiacal light, emission from the Milky Way, stray light, and thermal emission from the observatory.\footnote{\url{https://jwst-docs.stsci.edu/jwst-general-support/jwst-background-model}} In the wavelength range where NIRCam operates, this model has been found to be fairly consistent with early JWST data, for $\lambda \gtrsim 1.5 \mu m$. At shorter wavelengths, there are indications that the JWST background model could be somewhat overestimated, possibly due to limitations in the zodiacal light and stray light components \citep[e.g.,][]{Rigby23}.

This behavior is similar to what we observe in our WLM data. The sky brightness predicted by the ETC for our F150W, F250M, and F430M frames matches the ERS observed sky values, which we take to be the mode of the pixel brightness distribution across all detectors and exposures we acquired for that filter, to within  within 0.03 MJy/sr. However, for the F090W band, the ETC sky model predicts a brightness of 0.45 MJy/sr, while we measure a value of 0.28 MJy/sr. This is in line with what is reported by \citet{Rigby23}.

If the background model employed by the ETC consistently overestimates the sky brightness in the bluest JWST filters, this should result in the ETC systematically underestimating SNRs. For the faint sources in our WLM dataset, we estimate this effect to be of approximately 15\%, for the F090W band. We recommend that users consider this effect for observations at $\lambda \lesssim 1.5 \mu m$. However, we note that there are already active efforts to improve the JWST background model using existing JWST data (e.g., JWST-AR-4695; PI: Windhorst). As these improvements are incorporated into in new ETC iterations, we expect improved SNR predictions at these wavelengths.

Finally, we note that there may be instances in which effects contribute to the local sky brightness in addition to sources included by the ETC model background. Some examples include stray light artifacts, such as wisps and claws \citep[e.g., ][]{Bagley23,Windhorst23}, or persistence from bright sources \citep[e.g., ][]{Leisenring16}. In fact, this is the case for our \dracoii\ images, in which persistence from previous observations of Solar System objects resulted in prominent and irregular background features (see Figure~10 from \citealt{Weisz24}). This artificially increased our sky level by as much as 50\% in F090W. While such example is likely to be an extreme case, also due to the particularly low F090W sky at the date and coordinates of our \dracoii\ observations, it serves to illustrate that such events can occur and  can have meaningful impact on photometric performance. Another instance where the local background could be brighter than expectations is in observations of high-surface brightness objects, for which the unresolved light from the target itself can become a major contributor to the local sky brightness. This is the case of our M92 observations and it is further detailed in \S~\ref{sec:Crowding}.

\begin{figure*}
    \subfloat
        {\includegraphics[width=0.5\textwidth]{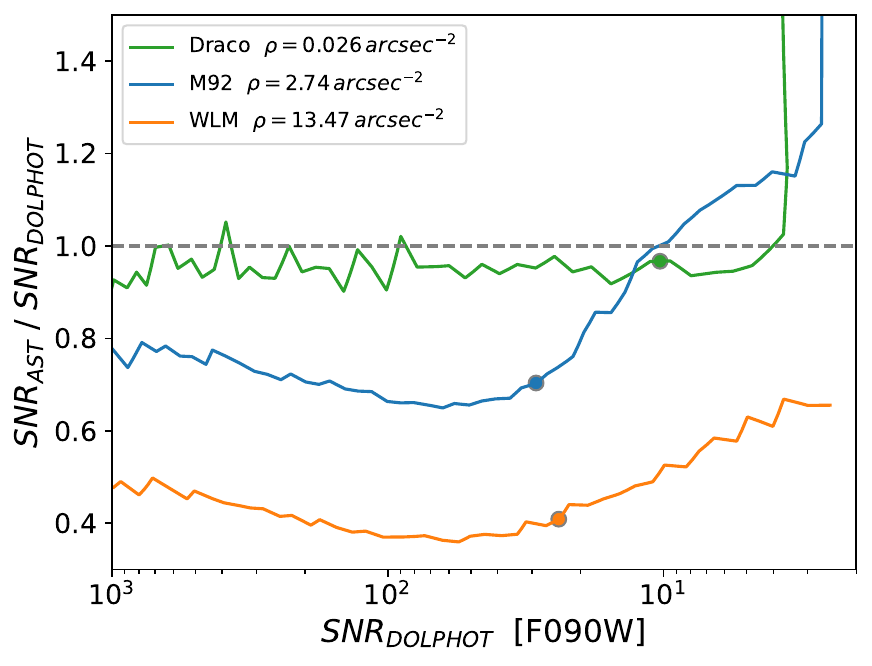}} \quad
    \subfloat
        {\includegraphics[width=0.5\textwidth]{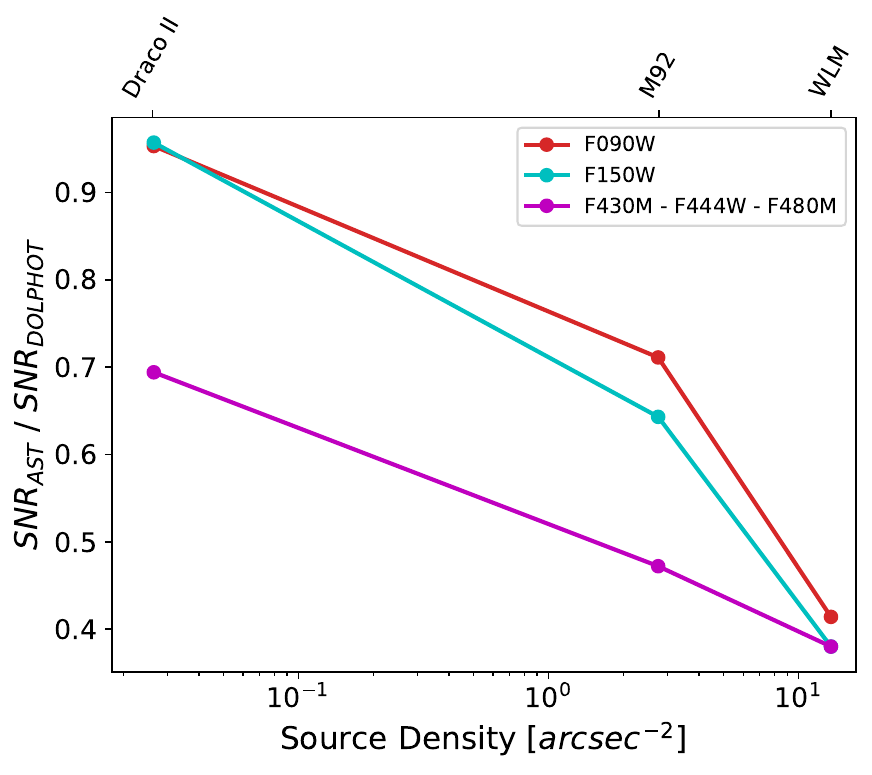}}  \quad

    \caption{Left: Ratio between the AST-based SNR and the DOLPHOT SNR, for the F090W filter, in Draco II (green), M92 (blue), and WLM (orange), as a function of DOLPHOT SNR. For each target, the SNR corresponding to 75\% completeness is marked by the circle. Right: Average Ratio between the AST-based SNR and the DOLPHOT SNR, as a function of observed source density, for F090W (red), F150W (cyan) and F430M/F444W/F480M (magenta).  For each data point, the corresponding target is reported on the top axis.}
    \label{fig:AST}
 
\end{figure*}

\subsection{The Impact of Crowding}
\label{sec:Crowding}
The analysis presented so far assumes that photometry is performed on an isolated point source. This is a reasonable approximation for certain observations (e.g., high redshift galaxies, sparse stellar fields, or shallow imaging). However, many observations of nearby stellar systems will be characterized by a meaningful degree of crowding \citep[e.g.,][]{Stetson77,Stetson88,Stetson94,Tolstoy98,Grebel00,Piotto02,Benjamin03,Zoccali03,Zaritsky04,Rejkuba05,Sarajedini07,Dalcanton09,Monachesi11,Williams14}. In those regimes, PSF-fitting photometry is the only viable approach and, even then, contamination introduced by nearby stellar sources becomes a major factor affecting photometric performance. 

A reliable estimation of photometric uncertainties in crowded fields can be achieved through the AST approach. By using the $\sim 3\times10^6$ ASTs, per target, produced by our program, we can therefore compare the real photometric precision achieved by our observations, with respect to the idealized estimate made for isolated point sources. As our three targets have significantly different crowding properties, we can also explore the importance of this effect as a function of stellar density.

We use the SNR values reported by DOLPHOT as a baseline against which to compare the AST-based SNRs. This choice ensures that all other factors contributing to the error budget (e.g., sky brightness, photometric aperture, assumptions about the exposure time) are the same and the effect of crowding is properly isolated in this test. We consider only SNR$<1000$, as this is the highest SNR we can reliably measure in the ASTs, due to numerical precision. We also estimate the average stellar density in our fields by counting the number of observed \textit{bona fide} stars (i.e., those that pass the \citealt{Warfield23} selection and have SNR $\ge5$ in the SW filters) in our catalogs and normalizing by the area of the NIRCam field of view (approximately 9.1 $arcmin^2$, for the 8 SW detectors). We use the number of detected sources because the importance of crowding does not depend only on the intrinsic stellar density but also on the photometric depth of the observations.

\begin{figure*}
        \includegraphics[width=\textwidth]{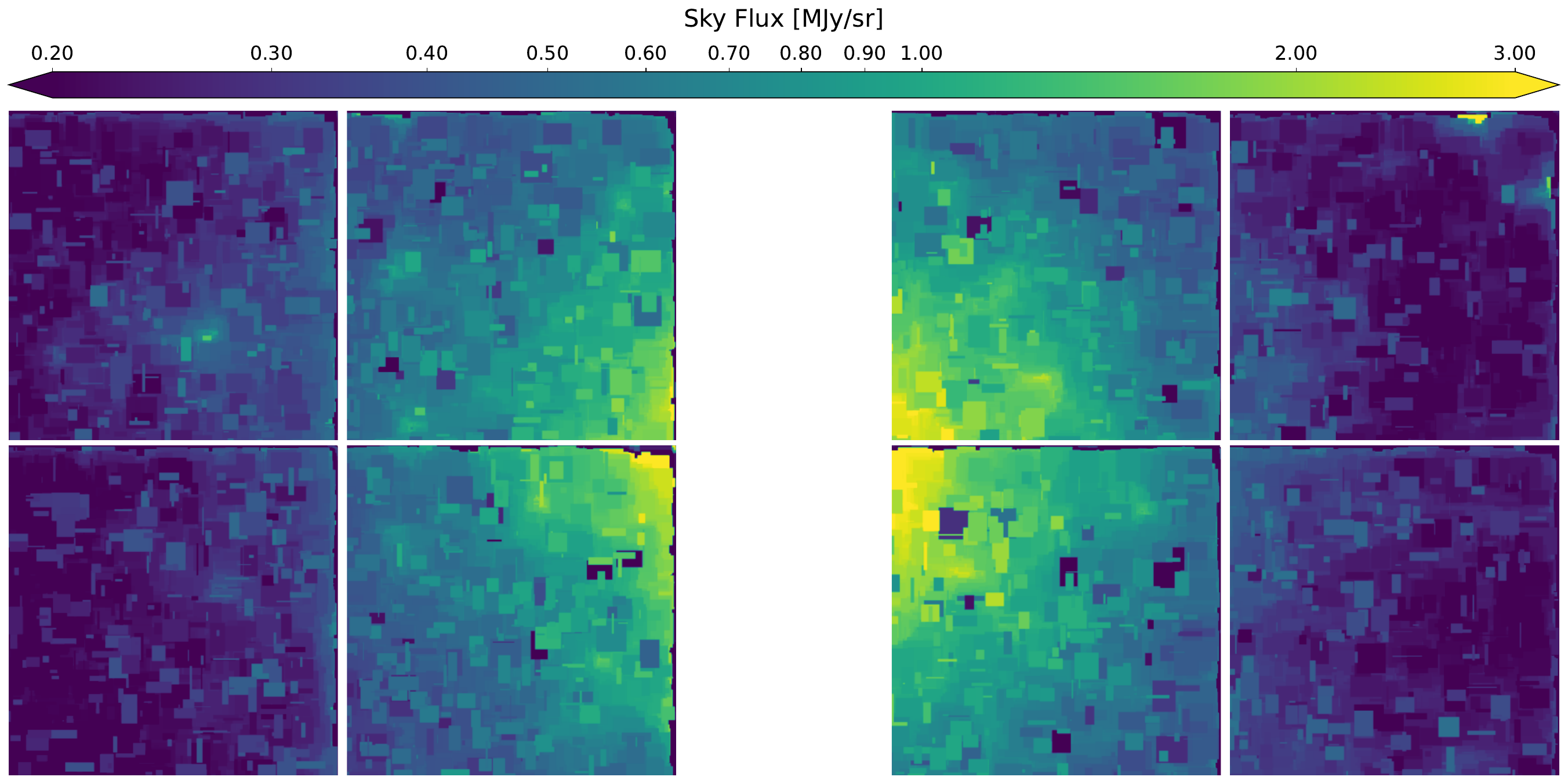}

    \caption{Map of the F090W sky brightness in our M92 NIRCam images. The sky flux has been calculated as the mode of the pixel brightness distribution over a 200~px~$\times$~200~px moving box, after subtracting the light from all detected point sources.}
    \label{fig:M92Sky}
 
\end{figure*}

The left panel of Figure~\ref{fig:AST} shows the ratio between the AST-based SNR in F090W and the DOLPHOT photon-based SNR, for the three ERS targets. In the case of \dracoii, the two SNRs are in excellent agreement (within 5\%) over the entire SNR range. The sharp increase at very low-SNR is an artifact caused by ASTs near the detection threshold, whose recovered photometric properties are very biased and do not lead to a meaningful SNR estimate. The almost perfect agreement between the SNR captured by the ASTs and the isolated-source calculation provided by DOLPHOT is a reflection of the very low stellar density of the \dracoii\ field.

As we move to higher stellar densities, with the two other ERS targets, we see that crowding becomes increasingly important, and the AST-based SNR is lower than what is reported by DOLPHOT. Additionally, the SNR ratio is not constant but rather exhibits a few notable features. Starting from the brightest sources ($SNR\sim1000$), the discrepancy between DOLPHOT and the ASTs increases with decreasing SNR. This is because fainter stars are more affected by crowding than brighter stars, as the stellar density increases with increasing magnitude.

\begin{figure*}
\centering
        \includegraphics[width=\textwidth]{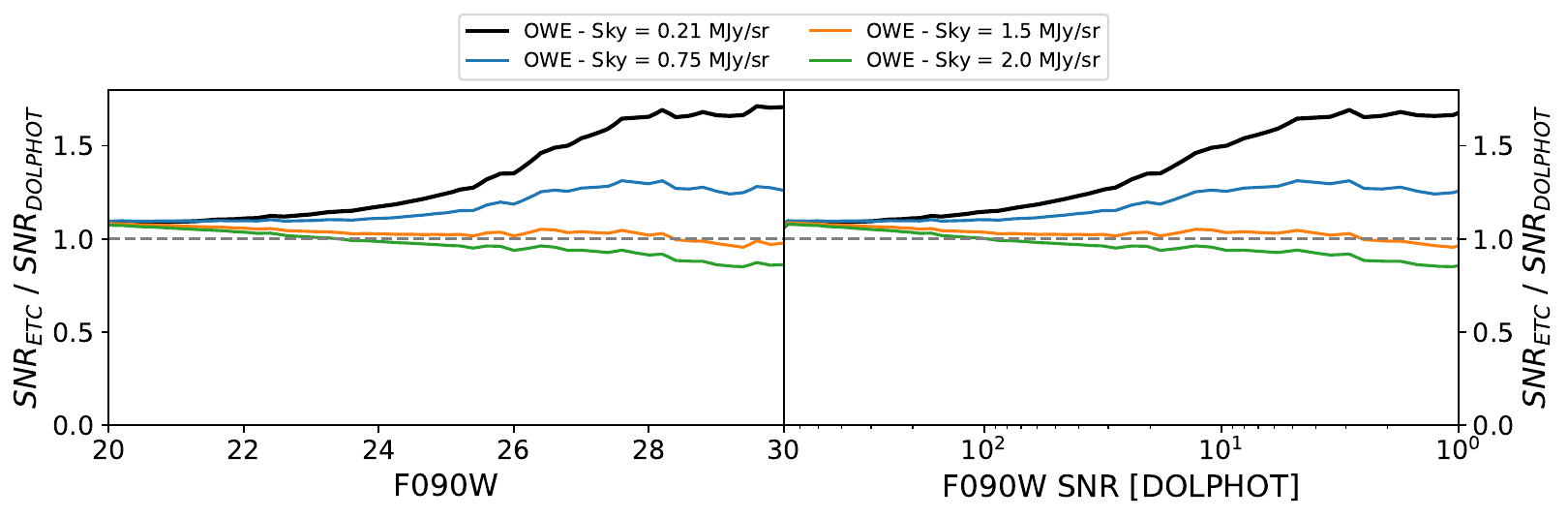}
    \caption{Ratio between the ETC SNR prediction and the DOLPHOT SNR, for the F090W photometry of our M92 observations, as a function of magnitude (left panel) and SNR (right panel). The ETC SNR is calculated using the OWE approach, with both flat-field and cosmic-ray effects disabled. The black line has been calculated using the ETC background model for the date and coordinates of our observations, resulting in a background brightness of 0.21 MJy/sr. The colored lines have been obtained by setting the ETC background brightness to 0.75 MJy/sr (blue), 1.5 MJy/sr (orange), and 2.0 MJy/sr (green), which are typical local sky values for the majority of our M92 sources.}
    \label{fig:Ratio_M92}
 
\end{figure*}

Eventually, the SNR ratio plateaus and then increases again (SNR $\lesssim 50$). This is a spurious effect caused by incompleteness. As the completeness level drops, artificial stars in the tails of the error distribution are more likely to be undetected (or discarded during reduction, or culling of the catalogs), compared to stars with good photometry. This narrows the error distribution and leads to artificially higher SNRs. For illustrative purposes, we report the SNR at which completeness reaches 75\% (marked by the circle). In M92 and WLM, this roughly corresponds to the inflection point in the SNR ratio.

As a way to quantify the effect of crowding, we calculate the average SNR ratio between SNR=1000 and the SNR at 75\% completeness. We show this in the right panel of Figure~\ref{fig:AST} for different filters, as a function of detected source density. All three ERS targets have been observed in the F090W and F150W filters, which allows us to trace their sensitivity to crowding as the density increases. On the other hand, the LW channel filters are heterogeneous. However, F430M, F444W, and F480M have very similar PSF sizes (0.144\arcsec-0.164\arcsec\ FWHM), so are expected to behave similarly.

From Figure~\ref{fig:AST}, we see how the effect of crowding increases with both source density and PSF size. For the F090W and F150W filters, the \dracoii\ field is essentially crowding-free, and the AST-based SNR is virtually identical to the isolated point-source estimate made by DOLPHOT. For these filters, the reported source density translates to an average distance to the nearest neighbor of roughly 100 times the PSF FWHM. Crowding begins to manifest in the significantly denser field of M92 (average neighbor distance of roughly 5-10 PSF FWHMs), and the AST-based SNR is only 60-70\% of the DOLPHOT estimate. This figure drops to approximately 40\% in the even denser field of WLM (average neighbor distance of 3-4 PSF FWHMs). The trend for the LW filters is similar, but the larger PSF results in a higher sensitivity to crowding. Even in the \dracoii\  field we see an appreciable reduction in the AST-based SNR, compared to the isolated source estimate. For reference, the average neighbor distance we measure in \dracoii\ is $\sim20\times$ the FWHM of the F480M PSF.

Figure~\ref{fig:AST} provides rough guidelines to approximate the effect of crowding on the ETC SNR estimates, provided a reasonable guess on the density of \textit{observed} sources, which depends on the intrinsic stellar density and on the desired photometric depth. However, this should only be intended as a rule of thumb, to gauge the importance of crowding effects on the final photometric performance. The relatively simple analysis presented here, the limited number of filters considered, and the small target sample of our ERS dataset are not enough to enable a precise prediction of the photometric performance. Furthermore, the detailed impact of crowding effects is likely to be influenced by the target luminosity function, the source spatial distribution, the use of multi-band data during reduction, and other program-dependent characteristics. For these reasons, observers who want a more rigorous assessment of photometric performance in crowded fields may wish to use a more sophisticated approach, such as producing and analyzing mock JWST images of their intended target (e.g., through the use of the Mirage software package\footnote{\url{https://www.stsci.edu/jwst/science-planning/proposal-planning-toolbox/mirage}}).

\begin{figure*}
        \includegraphics[width=\textwidth]{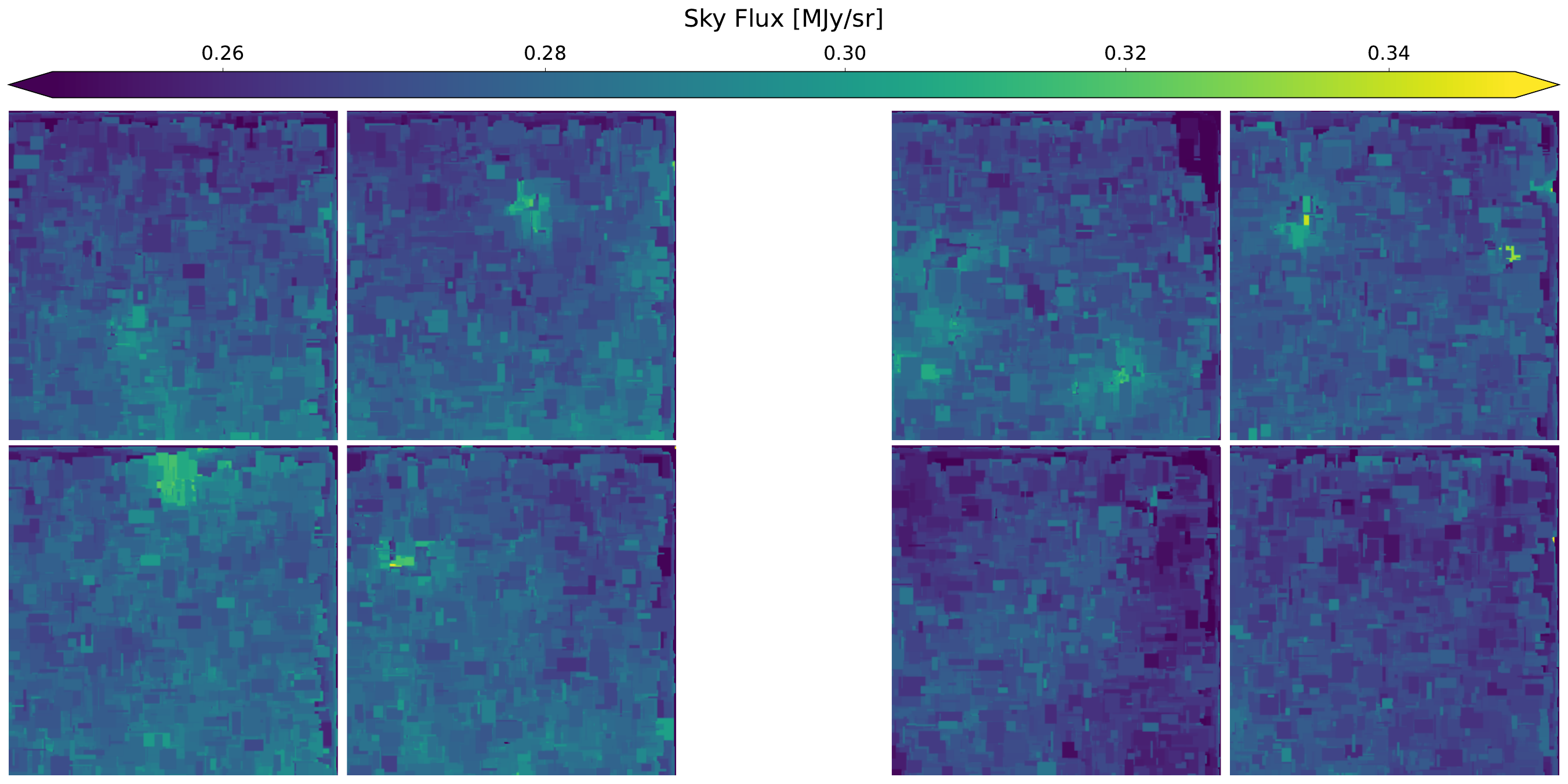}

    \caption{Same as Figure~\ref{fig:M92Sky}, but for the F090W images of the WLM field. Note the different dynamic range, compared to M92.}
    \label{fig:WLMSky}
 
\end{figure*}

A final complication that may arise when designing observations of high-surface brightness objects has to do with the impact of unresolved light from the target itself. This can arise from faint undetected stars or from the PSF wings of bright resolved stars, and can potentially add a significant contribution to the local background level, affecting the photometric SNR in a way that is not captured by the ETC sky model.

We clearly see this effect in our M92 observations. Figure~\ref{fig:M92Sky} shows a map of the measured F090W sky flux in our M92 NIRCam field. We calculate the sky brightness as the mode of the pixel brightness distribution over a 200~px~$\times$~200~px moving box. We measure the sky brightness in the \textit{residual} frames produced by DOLPHOT, i.e., the images in which the model light from the detected point sources has been subtracted. From Figure~\ref{fig:M92Sky}, a clear gradient is visible in the background flux, indicating that the unresolved light from the target itself is a significant contributor to the local sky brightness.

Accordingly, Figure~\ref{fig:Ratio_M92} shows that the photon-based SNR reported by DOLPHOT for M92 sources is significantly lower (roughly 60\%) than what is predicted by the ETC (using OWE estimates). The agreement with the ETC is mostly restored if the ETC background flux is manually adjusted to the range of values measured in our M92 frames.

We stress that this effect is expected to be dependent on the surface brightness of the target (more specifically, on the amount of flux per unit area that does not get resolved into individual stars), which does not necessarily correlate with the crowding properties of the frame, i.e., the average distance between neighboring point sources with respect to the PSF size. An observed field might have high surface brightness but low crowding (e.g., shallow imaging of a nearby dense stellar system) and a crowded field might belong to a low-surface brightness object (e.g., deep imaging of a distant, diffuse galaxy).

Indeed, we observe excess sky brightness in M92 \citep[$17\, {\rm mag/arcsec^2}\lesssim\mu_V\lesssim21 \, {\rm mag/arcsec^2}$, for the radii probed by our field,][]{Trager95} but not in WLM \citep[$\mu_V=23.7\, {\rm mag/arcsec^2}$,][Figure~\ref{fig:WLMSky}]{McConnachie12}, in spite of the much higher source density of the WLM field. Within the range of resolved stellar systems accessible by JWST, we therefore expect this to be a significant issue only in particularly dense objects, such as bright globular clusters, compact spheroidal galaxies like NGC205, or the inner regions of M31. For designing observations of such objects, analyzing mock JWST images is again the recommended strategy to estimate realistic photometric depths.

\section{Conclusions}
\label{sec:Conclusions}
In this paper, we leveraged the large catalogs of point-source photometric measurements, produced as part of the ERS-1334 program, to compare SNR estimates from the JWST ETC to the observed on-sky performance, over a large range of magnitudes, filters, and crowding regimes. We provide a brief summary of the major takeaways of this analysis:

\vspace{2mm}
i) For bright point sources, the SNR estimate from the ETC is in very good agreement with our empirical measurements if only photon shot noise is considered. However, current flat-field uncertainties dominate the error budget in the high-SNR regime, effectively setting a floor to the maximum achievable photometric precision. Observers interested in very high-SNR photometry can use a large number of dithers and wide photometric apertures to mitigate this effect.

\vspace{2mm}
ii) The full-saturation magnitude predicted by the ETC is compatible with what is determined from our dataset, within roughly 0.5 mag. However, the magnitude at which saturation is achieved varies with the precise source position within the central pixel. Observations taken with a modest amount of sub-pixel dithers (4-6) are expected to deliver brighter saturation limits, compared to the default ETC source positioning, over a reasonably complete sample of bright stars. The gain in the saturation magnitude is modest at the reddest wavelengths, but it can be as high as 1~mag for the bluest NIRCam filters.

\vspace{2mm}
iii) For faint point-sources, for which the sky is a significant source of uncertainty, the choice of extraction aperture in the ETC is a critical element in determining realistic SNR estimates, especially if PSF-photometry is meant to be used on the real data. The ideal aperture size is generally smaller that the ETC default value, varying as a function of filter and, to a lesser extent, observing conditions (source magnitude, sky brightness, exposure time). In the most extreme cases (i.e., faint sources observed with blue filters), using the ideal extraction aperture can lead to as much as a five fold reduction in the exposure time required to achieve a given SNR.

\vspace{2mm}
iv) While currently not implemented in the ETC, the addition of an optimally-weighted extraction (OWE) based SNR to the ETC outputs would provide significant benefits for the design of point-source observations. Using an OWE approach results in a better approximation of PSF-photometry performance at all magnitudes. It also simplifies the observation design by providing an extraction set-up that performs optimally with all filters and observing conditions. Calculating OWE-based SNRs is computationally inexpensive and can be readily done from the 2D simulation products already generated by Pandeia.

\vspace{2mm}
v) If the observed field has a meaningful degree of crowding, additional considerations are required to obtain realistic SNR forecasts. By using AST experiments, we quantify the impact of crowding on the effective SNR achieved by our observations. Our results provide coarse guidelines to approximate the effect of crowding on the ETC SNR estimates. Observers requiring a more detailed estimate of crowding effects, especially in high-surface brightness objects, are encouraged to produce and analyze mock JWST images of their intended target.

\section*{Acknowledgments}
We thank the anonymous reviewer for providing useful feedback and helping us improving this paper. This work is based on observations made with the NASA/ESA/CSA James Webb Space Telescope. The data were obtained from the Mikulski Archive for Space Telescopes at the Space Telescope Science Institute, which is operated by the Association of Universities for Research in Astronomy, Inc., under NASA contract NAS 5-03127 for JWST. These observations are associated with program DD-ERS-1334. The specific observations analyzed can be accessed via \dataset[doi: 10.17909/cn6n-xg90]{https://doi.org/10.17909/cn6n-xg90}

%

\vspace{5mm}
\facilities{JWST(NIRCam)}


\software{ \texttt{Astropy} \citep{Astropy}, \texttt{DOLPHOT} \citep{Dolphin16}, \texttt{IPython} \citep{IPython}, \texttt{Matplotlib} \citep{Matplotlib}, \texttt{NumPy} \citep{Numpy}, \texttt{Pandas} \citep{Pandas},Pandeia \citep{Pandeia}, and \texttt{SciPy} \citep{Scipy}
          }




\bibliography{sample631}{}
\bibliographystyle{aasjournal}



\end{document}